\documentclass[aps,prd,amsmath,11pt,nofootinbib,superscriptaddress,
floatfix]{revtex4}
\usepackage{graphicx}
\usepackage{bm}
\usepackage{times}
\usepackage{color}
\usepackage{amsfonts,amssymb,latexsym,amsmath}
\usepackage{epstopdf}

\newcommand{\itp}{\affiliation{CAS Key Laboratory of Theoretical Physics,
            Institute of Theoretical Physics,\\ Chinese Academy of Sciences,
            Beijing 100190, China}}
\newcommand{\bonn}{\affiliation{Helmholtz-Institut f\"ur Strahlen- und
             Kernphysik and Bethe Center for Theoretical Physics,\\
             Universit\"at Bonn,  D-53115 Bonn, Germany}}
\newcommand{\fzj}{\affiliation{Institute for
           Advanced Simulation, Institut f\"ur Kernphysik and
           J\"ulich Center for Hadron Physics,\\ 
           Forschungszentrum J\"ulich, D-52425 J\"ulich, Germany}}
\newcommand{\ific}{\affiliation{Instituto de F\'isica Corpuscular (IFIC),
             Centro Mixto CSIC-Universidad de Valencia,
             Institutos de Investigaci\'on de Paterna,
             Aptd. 22085, E-46071 Valencia, Spain}}
\newcommand{\ucas}{\affiliation{School of Physical Sciences,
            University of Chinese Academy of Sciences,
            Beijing 100049, China}}      
\newcommand{\mercia}{\affiliation{Departamento de F\'\i sica, 
           Universidad de Murcia, E-30071 Murcia, Spain}}

\begin{document}
\title{
Note on $\bm{X(3872)}$ production at hadron colliders and its molecular
structure }

\author{Miguel~Albaladejo} \mercia

\author{Feng-Kun~Guo} \itp\ucas

\author{Christoph~Hanhart} \fzj

\author{Ulf-G.~Mei{\ss}ner} \bonn\fzj

\author{Juan~Nieves} \ific

\author{Andreas~Nogga} \fzj

\author{Zhi Yang} \bonn


\begin{abstract}

  The production of the $X(3872)$ as a hadronic molecule in hadron
colliders is clarified. We show that the conclusion of Bignamini {\it et al}.,
Phys. Rev. Lett. {\bf 103} (2009) 162001, that the production of the $X(3872)$
at high $p_T$ implies a non-molecular structure, does not hold.
 In particular, using the well understood properties of the deuteron wave
 function as an example, we identify the relevant scales in the production
 process.
\end{abstract}

\maketitle

A widespread argument against the interpretation of the $X(3872)$ as a hadronic 
molecule is its copious production at hadron colliders:
Based on the inequality\footnote{Throughout the paper, 
the required charge 
conjugation combinations are implicit.}
\begin{eqnarray}
\sigma(\bar pp\to X) &\sim& \left| \int d^3{\mathbf k}\langle X|D^0\bar D^{*0}
({\mathbf k})\rangle\langle D^0\bar D^{*0}({\mathbf k})|\bar pp\rangle\right|^2
\nonumber \\
&\simeq& \left| \int_{\cal R} d^3{\mathbf k}\langle X|D^0\bar D^{*0}({\mathbf
k}) \rangle\langle D^0\bar D^{*0}({\mathbf k})|\bar pp\rangle\right|^2 \nonumber
\\
&\leq& \int_{\cal R} d^3 {\mathbf k} \left|\Psi({\mathbf k})\right|^2
\int_{\cal R} d^3 {\mathbf k}\left|\langle D^0\bar D^{*0}({\mathbf k})|\bar
pp\rangle\right|^2 \nonumber \\
 &\leq& \int_{\cal R} d^3 {\mathbf k}\left|\langle D^0\bar D^{*0}({\mathbf
k})|\bar pp\rangle\right|^2  \, 
\label{keyargument}
\end{eqnarray}
in Ref.~\cite{Bignamini:2009sk} it is claimed  that 
the fact that the $X(3872)$ was observed at high $p_T$ at Tevatron in $\bar pp$ 
collisions is inconsistent with the interpretation of $X(3872)$ as a $D^0\bar 
D^{0\, *}+c.c.$ molecule.
The whole argument depends crucially on the value of $\cal R$ which
specifies the region where the bound state wave function ``$\Psi(\mathbf k)$ 
is significantly different from zero''~\cite{Bignamini:2009sk}. In particular,
if $\cal R$ is chosen too small, the second line in Eq.~\eqref{keyargument}
is significantly smaller than the first and the whole argument is invalidated. 
Using a Gaussian wave function in 
Ref.~\cite{Bignamini:2009sk} a value ${\cal R}\simeq 35$~MeV of 
the order of the binding momentum was favored.
The so-estimated upper bound, 0.071~nb, is three 
orders  of magnitude smaller than the CDF measurement~\cite{Bauer:2004bc}, see 
Table~\ref{tab:crossec}. The authors concluded that the empirical
production rate at high $p_T$ is inconsistent with a prominent
 $D^0\bar D^{*0}$ molecular nature of the $X(3872)$. 
In this short note we 
challenge the reasoning of Ref.~\cite{Bignamini:2009sk} by showing that 
$\cal R$ should be significantly larger which leads to cross section estimates 
for the production of a molecular $X(3872)$ that are consistent with both the 
CDF and CMS measurements.

\begin{figure}[tbh]
\begin{center}
\includegraphics[width=0.45\linewidth]{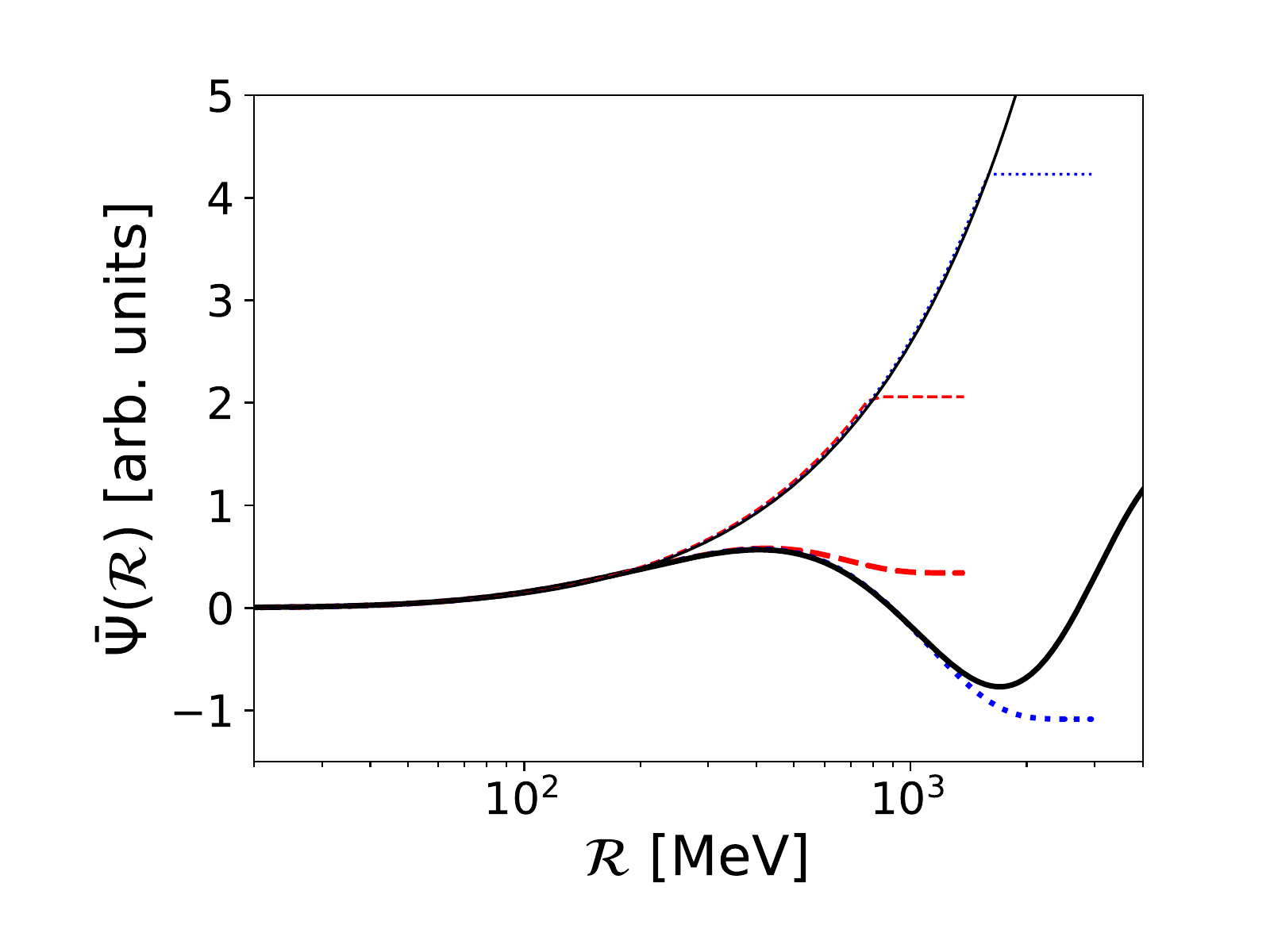}
\includegraphics[width=0.45\linewidth]{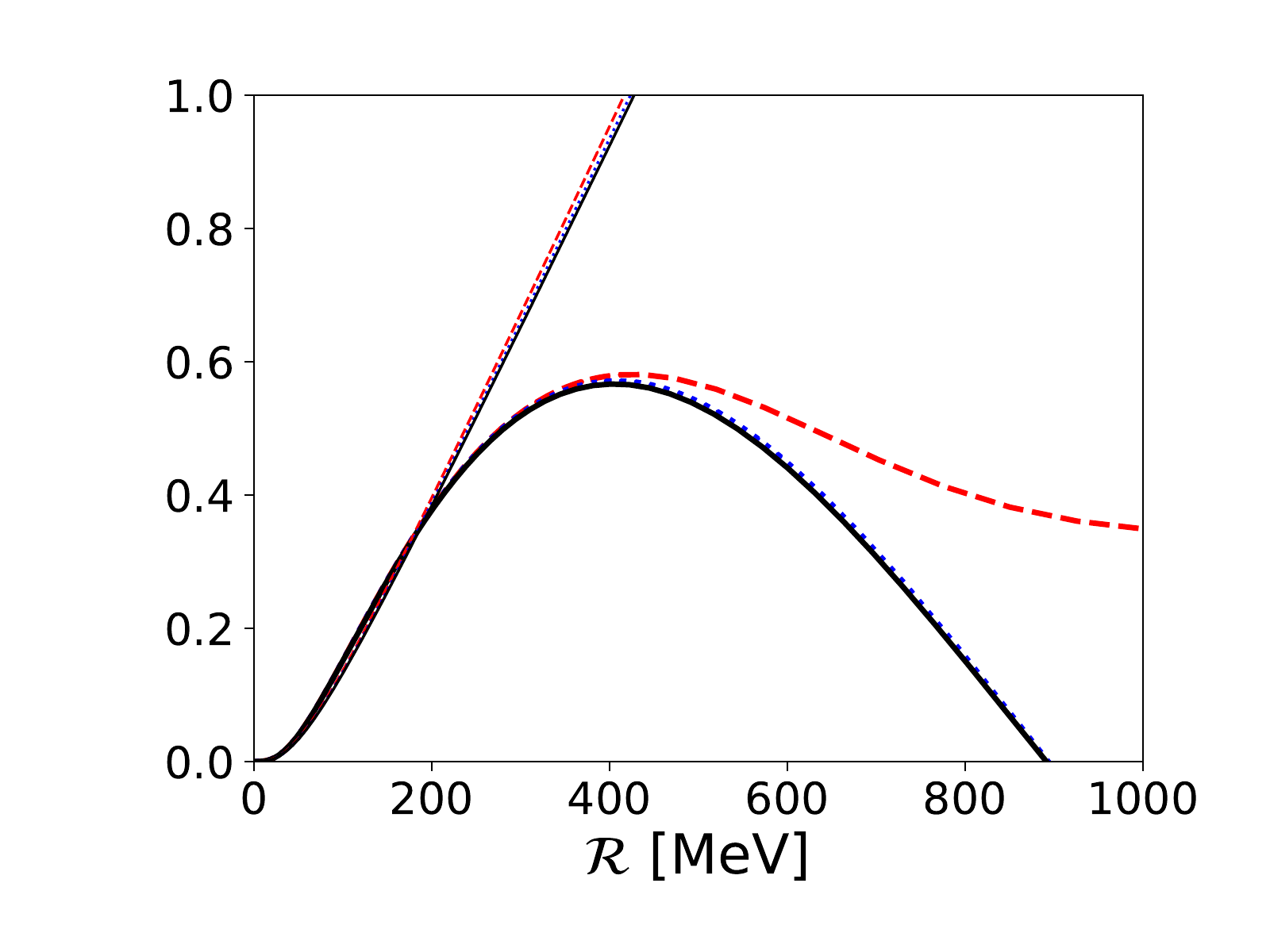}
\caption{Estimate of $\bar \Psi_\lambda({\cal R})$
for various deuteron wave functions: Results for $\lambda = 0.8, 1.6, 
4$~GeV are shown as red (dashed), blue (dotted) and black (solid) curves, respectively. 
The thick (thin) lines depict
the results with (without) OPE. 
The right panel is a zoom in a linear scale to the relevant $\cal R$
region.}
\label{fig}
\vspace{-3mm}
\end{center}
\end{figure}

In order to quantify which values of $\cal R$ are appropriate to justify 
the approximation,
we may assume that the production amplitude of $D\bar 
D^*$ pairs from $\bar pp$ collisions is (almost) a
constant~\cite{Artoisenet:2009wk}. Then we are left with the wave function 
averaged in the $\cal R$  region: $\bar 
\Psi_\lambda({\cal R}) \equiv \int_{\cal R} d^3 {\mathbf k} \,
\Psi_\lambda({\mathbf k})$,
where $\lambda$ specifies a regulator that needs to be introduced
to render the wave function well defined.
To understand which value of ${\cal R}$ is appropriate to 
largely saturate $\bar \Psi_\lambda({\cal R})$, we now switch to the
deuteron, a well established molecular proton-neutron bound state
with a binding momentum $\gamma_d\simeq45$~MeV.
In the case of the deuteron, as well as in that of the
$X(3872)$,  the range of forces is
of ${\cal O}(M_\pi^{-1})$, the pion Compton wave length.
Figure~\ref{fig} shows $\bar 
\Psi_\lambda({\cal R})$ for the deuteron with two sets of wave functions, one
generated from a potential 
with a short-ranged term and one-pion exchange (OPE) and 
the other without OPE~\cite{Nogga:2005fv}. 
For the former a Gaussian regulator is used and the small 
$D$-wave component is not shown. For the latter a sharp momentum space
cut off is used for simplicity.
One sees that $\bar \Psi_\lambda({\cal R})$ is far
from being saturated for ${\cal R}\simeq\gamma_d$ for all values of 
$\lambda$. A much larger value of 
${\cal R}\sim$300~MeV$\sim 2M_\pi$ needs to be taken for the second line in
Eq.~\eqref{keyargument} to be a good approximation of the first. The same 
should also be true for the $X(3872)$ as a $D\bar D^*$ 
molecule, since the range of forces is the same.
A similar value was also favored in
Ref.~\cite{Artoisenet:2009wk} based on rescattering arguments. In fact, 
Ref.~\cite{Bignamini:2009sk} noted that with such a value of $\cal R$ the 
upper bound becomes consistent with the CDF result. 


\begin{figure}[tbh]
\begin{center}
\includegraphics[width=0.5\linewidth]{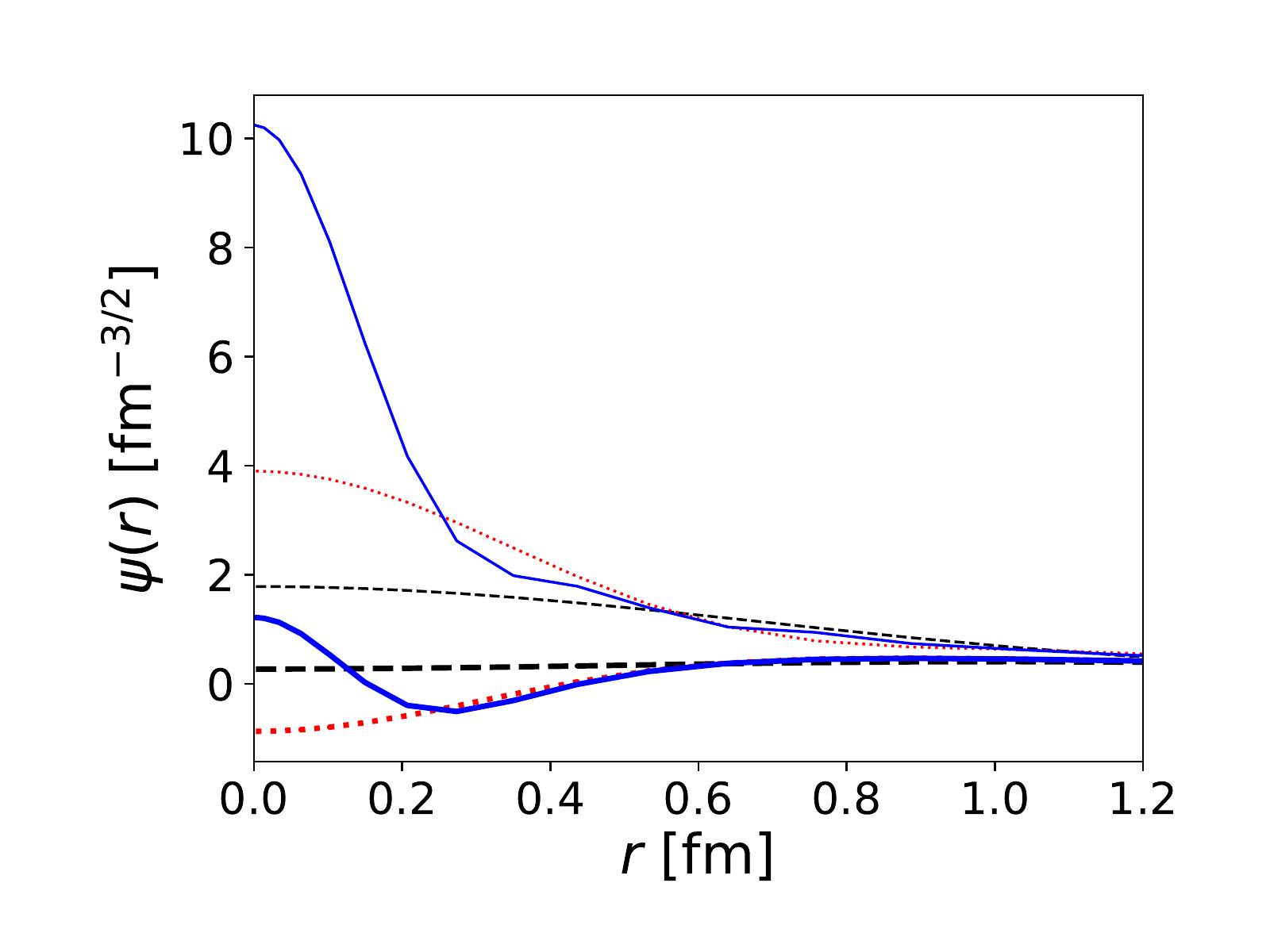}
\caption{The $S$-wave wave function of the deuteron calculated using an OPE 
potential with a cutoff $\lambda$~\cite{Nogga:2005fv}. Results for $\lambda = 
0.8, 1.6$ and 4~GeV are shown as red (dashed), blue (dotted) and black (solid) curves, respectively. 
The thick (thin) lines depict the results with (without) OPE. The wave functions have been normalized 
so that the magnitude of the tails agree.  
}
\label{fig2}
\vspace{-3mm}
\end{center}
\end{figure}

At the same time Fig.~\ref{fig}  illustrates that, within a hadronic theory, the
wave function at short distances is not a well defined object --- it is scheme
and regulator dependent. This can be seen from the $S$-wave deuteron wave
function calculated using different potentials as well as  different values  for
the regulator shown in Fig.~\ref{fig2} (for more details, we refer to
Ref.~\cite{Nogga:2005fv}).
Although the wave functions are very different at short distances, their
long-distance tails are universal. In fact the importance of this tail is a
direct measure of the molecular admixture in the wave function --- for a recent
review on hadronic molecules
 see Ref.~\cite{Guo:2017jvc}. A production process sensitive to such tails
is, e.g., the reaction $Y(4260)\to \gamma X(3872)$ --- because of this feature
is was possible to predict the rate for this reaction~\cite{Guo:2013nza}
starting from the known rate for $Y(4260)\to \pi Z_c(3900)$  assuming that 
the $Y(4260)$  and  the $Z_c(3900)$ were a $D_1(2420)\bar D$ and an isovector
$D\bar D^*$ hadronic molecules, respectively, as proposed in
Ref.~\cite{Wang:2013cya}.

There are various reactions that are sensitive to the short
ranged structure of the wave function which, as demonstrated above,
is  regulator dependent --- in particular they are not 
sensitive to the molecular component. Since within a consistent field theory
observables cannot be sensitive to the regulator employed, there must
be an additional short ranged operator of unknown strength at the leading
order that absorbs the regulator dependence. As a result, contrary to
the transitions mentioned in the previous paragraph, the hadronic
theory is not predictive for those reactions. Example of those 
are not only radiative decays of $X(3872)$ to vector states~\cite{Guo:2014taa}
 but also the production reactions discussed in this note.
 
\begin{table}[t]
\begin{ruledtabular}
\caption{Estimated inclusive cross sections compared with the 
CDF~\cite{Bauer:2004bc} and CMS~\cite{Chatrchyan:2013cld} measurements
converted into cross sections~\cite{Guo:2014sca}.
Results outside (inside) brackets are obtained using Herwig (Pythia) 
with the same
kinematical cuts as in~\cite{Guo:2014sca}.}
\begin{tabular}{ l  c  c  c  c } 
  $\sigma(pp/\bar p\to X)$ [nb]  & 
Exp.  & $\Lambda$=0.1 GeV   & $\Lambda$=0.5 GeV & $\Lambda$=1.0 GeV \\
 \hline
 Tevatron   &  37-115   & 0.07 (0.05)          & ~~7 (5)        & 29 (20) \\%
 LHC-7      &  13-39    & 0.12 (0.04)          & 13 (4)        & 55 (15) \\%
\end{tabular}
\label{tab:crossec}
\end{ruledtabular}
\end{table}

However, at least in the theory with OPE the wave function at short
distances is bounded and one may use renormalisation group and naturalness 
arguments to claim
that its typical values may still be used to estimate the order-of-magnitude of
the corresponding contribution.
To be more quantitative, in Table~\ref{tab:crossec}, we present the cross 
section estimates for  inclusive $X(3872)$ production 
compared with the CDF~\cite{Bauer:2004bc}  and 
CMS~\cite{Chatrchyan:2013cld} measurements by computing the short-distance 
$D\bar D^*$ production with the Monte Carlo event generators 
Pythia~\cite{Sjostrand:2007gs} and Herwig~\cite{Bahr:2008pv} and the 
long-distance part using an effective field theory treating the $X(3872)$ as a 
hadronic molecule 
following~\cite{Artoisenet:2009wk,Artoisenet:2010uu,Guo:2014sca, Albaladejo:2015dsa}. A Gaussian 
regulator with a cutoff $\Lambda$, which roughly amounts to $2\sqrt{2/\pi}{\cal 
R}\simeq1.6{\cal R}$, is used. The results with $\Lambda=0.1$~GeV, 
with only neutral charmed mesons included, are much smaller than the data. 
However, the order-of-magnitude estimated cross sections become consistent with 
measurements when a larger cutoff of 
$[0.5,1.0]$~GeV, roughly corresponding to ${\cal R}\in[0.3,0.6]$~GeV, 
is used. Notice that for such a large cutoff, the $D^+D^{*-}$ channel with a
binding momentum of 126~MeV becomes dynamical and has been included. Its 
inclusion has led to an enhancement of about a factor of three. It is apparent 
that the order-of-magnitude estimates are consistent with the CDF and CMS 
measurements.

\medskip

\begin{acknowledgments}

This work is supported in part by DFG and NSFC through funds provided to the
Sino-German CRC 110 ``Symmetries and the Emergence of Structure in QCD" (NSFC
Grant No.~11621131001, DFG Grant No.~TRR110), by NSFC (Grant No.~11647601), by 
the CAS Key Research Program of Frontier Sciences (Grant No.~QYZDB-SSW-SYS013), 
by the Thousand Talents Plan for Young Professionals, by the CAS President's 
International Fellowship Initiative (PIFI) (Grant No.~2017VMA0025), and by 
Spanish Ministerio de Econom\'\i a y Competitividad and European FEDER under 
contracts FIS2014-51948-C2-1-P and SEV-2014-0398. Part of the computations have been 
performed on JUQUEEN and JURECA of the JSC, J\"ulich, Germany.   

\end{acknowledgments}


\begin{thebibliography}{99}

\bibitem{Bignamini:2009sk} 
  C.~Bignamini, B.~Grinstein, F.~Piccinini, A.~D.~Polosa and C.~Sabelli,
  Phys.\ Rev.\ Lett.\  {\bf 103}, 162001 (2009)
  [arXiv:0906.0882 [hep-ph]].

\bibitem{Bauer:2004bc} 
  G.~Bauer [CDF Collaboration],
  Int.\ J.\ Mod.\ Phys.\ A {\bf 20}, 3765 (2005)
  [hep-ex/0409052].

\bibitem{Artoisenet:2009wk} 
  P.~Artoisenet and E.~Braaten,
  Phys.\ Rev.\ D {\bf 81}, 114018 (2010)
  [arXiv:0911.2016 [hep-ph]].

\bibitem{Nogga:2005fv} 
  A.~Nogga and C.~Hanhart,
  Phys.\ Lett.\ B {\bf 634}, 210 (2006)
  [nucl-th/0511011].
  
\bibitem{Guo:2017jvc} 
  F.-K.~Guo, C.~Hanhart, U.-G.~Mei{\ss}ner, Q.~Wang, Q.~Zhao and B.-S.~Zou,
  arXiv:1705.00141 [hep-ph].

\bibitem{Chatrchyan:2013cld} 
  S.~Chatrchyan {\it et al.} [CMS Collaboration],
  JHEP {\bf 1304}, 154 (2013)
  [arXiv:1302.3968 [hep-ex]].

\bibitem{Sjostrand:2007gs} 
  T.~Sjostrand, S.~Mrenna and P.~Z.~Skands,
  Comput.\ Phys.\ Commun.\  {\bf 178}, 852 (2008)
  [arXiv:0710.3820 [hep-ph]].

\bibitem{Bahr:2008pv} 
  M.~Bahr {\it et al.},
  Eur.\ Phys.\ J.\ C {\bf 58}, 639 (2008)
  [arXiv:0803.0883 [hep-ph]].
  
\bibitem{Artoisenet:2010uu} 
  P.~Artoisenet and E.~Braaten,
  Phys.\ Rev.\ D {\bf 83}, 014019 (2011)
 [arXiv:1007.2868 [hep-ph]].

\bibitem{Guo:2014sca} 
  F.-K.~Guo, U.-G.~Mei{\ss}ner, W.~Wang and Z.~Yang,
  Eur.\ Phys.\ J.\ C {\bf 74}, 3063 (2014)
  [arXiv:1402.6236 [hep-ph]].
  
  
 \bibitem{Guo:2013nza} 
  F.-K.~Guo, C.~Hanhart, U.-G.~Mei{\ss}ner, Q.~Wang and Q.~Zhao,
  Phys.\ Lett.\ B {\bf 725}, 127 (2013)
  [arXiv:1306.3096 [hep-ph]].
    
  \bibitem{Wang:2013cya}
  Q.~Wang, C.~Hanhart and Q.~Zhao,
  Phys.\ Rev.\ Lett.\  {\bf 111}, 132003 (2013)
  [arXiv:1303.6355 [hep-ph]].


\bibitem{Guo:2014taa} 
  F.-K.~Guo, C.~Hanhart, Y.~S.~Kalashnikova, U.-G.~Mei{\ss}ner and
  A.~V.~Nefediev,
  Phys.\ Lett.\ B {\bf 742}, 394 (2015)
  [arXiv:1410.6712 [hep-ph]].

 \bibitem{Albaladejo:2015dsa}
  M.~Albaladejo, F.-K.~Guo, C.~Hidalgo-Duque, J.~Nieves and
  M.~P.~Valderrama,
  Eur.\ Phys.\ J.\ C {\bf 75},  547 (2015).
  [arXiv:1504.00861 [hep-ph]].
  
\end{thebibliography}
\end{document}